# Diffuse Molecular Clouds and the Molecular Interstellar Medium from $^{13}$CO Observations of M33

*Christine D. Wilson*

*Department of Physics and Astronomy, McMaster University, Hamilton, Ontario, Canada L8S 4M1*

*and*

*Constance E. Walker*

*Department of Astronomy, University of Texas and Steward Observatory,*

*University of Arizona, Tucson AZ U.S.A. 85721*

## Abstract

We have obtained $^{12}$CO and $^{13}$CO J=1-0 observations of the nearby spiral galaxy M33 to try to resolve the long-standing discrepancy between $^{12}$CO/$^{13}$CO line ratios measured in Galactic giant molecular clouds and external galaxies. Interferometer maps of the molecular cloud MC20 give a $^{12}$CO/$^{13}$CO line ratio of $7.5\pm2.1$, which agrees reasonably well with the line ratio measured in Galactic giant molecular clouds. In contrast, the $^{12}$CO/$^{13}$CO line ratio obtained from single dish data is $10.0 \pm 0.9$, significantly higher than Galactic values but in good agreement with line ratios measured in other galaxies. The interferometer map of MC20 reveals that the cloud has similar spatial and velocity extents in the two lines, and thus the high single dish line ratio cannot be due to different filling factors in the two lines. In addition, the single dish data show no evidence for significant variations in the line ratio with metallicity, which eliminates abundance changes as the explanation for the high single dish line ratio. We conclude that the high $^{12}$CO/$^{13}$CO line ratios observed in M33, and in the disks of spiral galaxies in general, are due to the presence of a population of diffuse molecular clouds. We obtain a lower limit to the $^{12}$CO/$^{13}$CO line ratio in the diffuse molecular clouds of $13 \pm 5$. The lower limit to the fraction of the total $^{12}$CO emission from M33 that originates in the diffuse clouds is $30\pm30\%$, while the upper limit is $\sim 60\%$. We have combined our line ratios with published measurements of the $^{12}$CO J=2-1 to J=1-0 line ratio to determine the $^{12}$CO and $^{13}$CO column densities to within an order of magnitude, but the density and the temperature of the gas are not well constrained by these measurements. We estimate that differences in the physical conditions in diffuse and giant molecular clouds are unlikely to cause the overall molecular gas mass calculated for the central kiloparsec of M33 to be in error by more than a factor of two.

## 1. Introduction

Observations of multiple transitions and isotopes of CO can be used to determine the physical properties of the molecular interstellar medium in external galaxies. Particularly useful are the lines of $^{13}$CO, since this emission is usually optically thin and hence a more direct tracer of optical depth and column density than the more abundant $^{12}$CO molecule. However, ever since the first observations of $^{13}$CO in external galaxies (Encrenaz et al. 1979), an interesting difference has been noted between giant molecular clouds (GMCs) in the Milky Way and molecular emission from external galaxies: the $^{12}$CO/$^{13}$CO line ratio is substantially smaller in Galactic giant molecular clouds (3-6) than in external galaxies (typically $> 10$). The larger line ratios observed in external galaxies have been attributed to different beam filling factors in the two lines (Encrenaz et al. 1979), different intrinsic line ratios in diffuse molecular clouds and giant molecular clouds (GMCs) (Polk et al. 1988; Young & Sanders 1986), and different [$^{12}$CO]/[$^{13}$CO] abundance ratios (Casoli, Dupraz, & Combes 1992). Indeed, each of these effects may be important in different galactic environments.

Most observations of the $^{12}$CO/$^{13}$CO line ratio in the Milky Way have concentrated on the peak emission regions of Galactic GMCs, with values ranging from 3 (Gordon & Burton 1976) to 6 (Solomon, Scoville, & Sanders 1979). However, Knapp & Bowers (1988) have measured the line ratio in five diffuse molecular clouds to be $21 \pm 8$. Using this result, Polk et al. (1988) estimate that diffuse molecular clouds produce 40-60% of the total $^{12}$CO emission from the Galaxy. In studies of external galaxies, which contain both diffuse and giant molecular clouds within the beam, the observed line ratio would fall naturally between these two extremes. Since the conversion factor between $^{12}$CO flux and $H_2$ column density may be quite different in the diffuse molecular clouds because of their lower column densities (de Vries, Heithausen, & Thaddeus 1987), a significant $^{12}$CO flux contribution from these clouds could complicate molecular gas mass determinations for external galaxies.

In nearby galaxies, interferometric observations of individual GMCs can determine whether GMCs have the same filling factor in both $^{12}$CO and $^{13}$CO. In addition, the effect of abundance variations can be estimated by observing regions with different metallicity within the same galaxy. Even with current millimeter-wave interferometers, the small size of diffuse molecular clouds prevents their direct study in external galaxies. However, interferometric observations of individual GMCs can be combined with single dish observations to constrain both the $^{12}$CO/$^{13}$CO line ratio in diffuse molecular clouds and the fraction of the total $^{12}$CO flux emitted by these clouds. In this paper, we present a detailed study of the $^{13}$CO J=1-0 emission in the Local Group spiral galaxy M33. The observations and data reduction are discussed in §2. The $^{12}$CO/$^{13}$CO line ratios obtained from the single dish and the interferometric observations are discussed in §3. The interpretation of the line ratio in terms of the three possible explanations is discussed in §4. The properties of diffuse molecular clouds in M33 are discussed in §5. The $^{12}$CO/$^{13}$CO line ratios are combined with published $^{12}$CO J=2-1 data to constrain the physical properties of the molecular gas using large velocity gradient models in §6. The paper is summarized in §7.

## 2. Observations and Data Reduction
## 2.1 Single Dish CO Observations

Seven positions in M33 centered on previously identified giant molecular clouds (Wilson & Scoville 1990) were observed with the NRAO 12 m telescope[1] in the $^{12}$CO and $^{13}$CO J=1-0 lines in 1991 October. The 256 channel 1 MHz filterbank was configured in parallel mode to detect both polarizations. Total integration times (on + off source) ranged from 12 to 36 minutes for the $^{12}$CO line and 102 to 162 minutes for the $^{13}$CO line. Small ($\sim 4$ MHz) frequency shifts were made during the $^{12}$CO and $^{13}$CO observations to minimize the chance of a spurious detection due to problems with individual channels of the filter banks. The weather during the observations was generally clear and stable. Exceptions occurred in observing MC32, for which the measured system temperatures occasionally varied on timescales of minutes due to fog, and in observing MC19, for which the observations were interrupted by an extended period of unstable system temperatures due to clouds. Except for these two periods of bad weather, single sideband system temperatures were 410-960 K for the $^{12}$CO observations and 220-420 K for the $^{13}$CO observations.

The pointing was checked frequently throughout the run (4-10 times per night). The pointing was very stable, with no strong dependence on azimuth or elevation, and the pointing accuracy is estimated to be $\leq 8''$. No observations were obtained within one hour of transit, when M33 is at very high elevation angles and the pointing and tracking of the telescope is less accurate. To provide a further check on the pointing stability, $^{12}$CO measurements were obtained both before and after the $^{13}$CO observations for all positions except MC32. For four of the positions (MC6-10, MC13, MC19, MC20), the $^{12}$CO line profiles observed before and after the $^{13}$CO observations agreed very well, with both the peak temperatures and integrated intensities agreeing to better than 15%. For MC1-5, the peak temperatures agreed well but the integrated intensities differed by 20% and the shapes of the spectra were quite different. For NGC 604, the integrated intensities agreed very well, but the line profiles were quite different. The shapes of the line profiles change quite rapidly with position around NGC 604 (Wilson & Scoville 1992) and near the nucleus of M33 and the difference between the two pairs of $^{12}$CO spectra may indicate slight pointing errors.

Four calibration sources (DR21(OH), W51, Orion A, and IRC 10216) were observed in $^{12}$CO throughout the run to assess the stability of the calibration; two of these sources were also observed in $^{13}$CO. The average peak temperatures quoted here are at 1 MHz resolution and are based on two measurements on different days for each transition, except for the $^{12}$CO average for DR21(OH), for which we have seven measurements, and W51, for which we have only a single measurement. The uncertainty quoted is the rms dispersion rather than the uncertainty in the mean. The $T_R^*$ temperature scale is used for the single dish data throughout this paper. (For reference, temperatures may be converted from the $T_R^*$ scale to the $T_{mb}$ scale using $T_{mb} = T_R^*/\eta_m^*$, where $\eta_m^* = 0.84$ at 110 GHz and $\eta_m^* = 0.82$ at 115 GHz (NRAO User's Manual).) For DR21(OH), the $^{12}$CO peak temperature is $30.0 \pm 1.6$ K while the $^{13}$CO peak temperature is $13.4 \pm 0.5$ K. For Orion A, the $^{12}$CO peak temperature is $80.6 \pm 4.0$ K, while the $^{13}$CO peak temperature is $11.6 \pm 0.6$ K. For IRC 10216, the peak

---

[1] The National Radio Astronomy Observatory is operated by Associated Universities, Inc., under cooperative agreement with the National Science Foundation.

$^{12}$CO temperature is 8.4 ± 0.6 K, while our single $^{12}$CO measurement of W51 yields a peak temperature of 39.4 K. Thus the temperatures of the calibrator sources are repeatable at the 5% level and we adopt 5% as the lower limit to the systematic uncertainty in our single dish calibration.

The data were reduced using the single dish data reduction package COMB. Scans obtained during periods of bad or unstable weather were rejected using the chart recorder trace of the receiver total power. The remaining scans from each receiver were inspected for obvious problems with the data. Three scans from one receiver were rejected, one $^{12}$CO scan on NGC 604 with a strong standing wave pattern, and two $^{13}$CO scans on MC1-5 with very large baseline slopes. No other potentially subjective selection of the data has been performed. The scans for each position were averaged together to produce the raw spectrum for each transition, which was then fitted with a first order baseline. The $^{13}$CO spectra were reinterpolated to have the same velocity resolution as the $^{12}$CO data (0.96 MHz = 2.6 km s$^{-1}$ at 110.2 GHz), while the $^{12}$CO data were left at the original sampling. The rms noise ($\sigma$), integrated intensity, and peak antenna temperature are given for each of the observed positions in Table 1 and the individual spectra are shown in Figure 1. The uncertainty in the integrated intensity is given by $2.6\sigma\sqrt{N_{line}}\sqrt{1 + N_{line}/N_{base}}$, where $N_{line}$ is the number of (2.6 km s$^{-1}$) channels over which the line was integrated and $N_{base}$ is the number of channels used in determining the baseline.

## 2.2 Interferometric CO Observations

One field centered on the giant molecular cloud MC20 was observed in the $^{13}$CO J=1-0 line with the Owens Valley Millimeter-Wave Interferometer between 1992 May and 1992 June. The field center was $\alpha(1950) = 01^h\ 31^m\ 10.9^s$, $\delta(1950) = 30^o\ 25'\ 24''$ and the data have a spectral resolution of 0.25 MHz. The field was observed in three configurations and typical single sideband system temperatures at the zenith were 280-350 K. Absolute flux calibration was obtained from observations of Uranus. The flux of the quasar 0133+476 obtained in the three tracks agreed to within 10% and thus we adopt 10% as a lower limit to the systematic uncertainty in our absolute calibration of these data.

$^{12}$CO J=1-0 observations of MC20 were obtained by Wilson & Scoville (1990) between 1987 December and 1988 May using the 32-channel 1 MHz filterbank. The field center was $\alpha(1950) = 01^h\ 31^m\ 10.5^s$, $\delta(1950) = 30^o\ 25'\ 20''$, slightly different from the field center used at $^{13}$CO. The field was observed for two partial tracks in each of three configurations and typical *double* sideband system temperatures at the zenith were 400-600 K. Absolute flux calibration was obtained from observations of Mars, 3C84, and 0133+476. The flux of the quasar 0133+476 obtained in the six tracks agreed to within 20% and thus we adopt 20% as a lower limit to the systematic uncertainty in our absolute calibration of these data.

Cleaned channel maps of both lines were made using the MIRIAD software package. Uniform weighting was used to obtain the highest possible spatial resolution and the restoring beam was 6.1″ × 7.8″. The field center of the $^{12}$CO data was shifted to correspond to the $^{13}$CO field center and the flux cutoff for cleaning was set to 0.28 Jy beam$^{-1}$ (1.5 times the rms noise in a single channel). The $^{13}$CO data were binned to the 2.6 km s$^{-1}$ velocity resolution of the $^{12}$CO data and the flux cutoff for cleaning was set to 0.086 Jy beam$^{-1}$ (1.5 times the rms noise). A gaussian beam with a width of 66″ (full-width half-maximum) was used to correct the fluxes for attenuation due to the primary beam. Integrated intensity maps of the $^{12}$CO and $^{13}$CO emission of MC20 were made by averaging the six channels that contain significant emission in both lines. The resulting dirty maps were cleaned using $1.5\sigma$ flux cutoffs and the same restoring beam as the channel maps.

### 3. The $^{12}$CO to $^{13}$CO J=1-0 Line Ratio in M33

We can combine our $^{12}$CO and $^{13}$CO interferometric observations of MC20 to measure the $^{12}$CO/$^{13}$CO J=1-0 line ratio, $R$. The integrated intensity line ratio is $R = 7.5 \pm 2.1$, where in addition to the measurement uncertainties we have included the uncertainty in the absolute calibration of the $^{12}$CO and $^{13}$CO data (see §2). The line ratio obtained from the peak brightness temperatures is $5.9 \pm 1.8$, again including both measurement and calibration uncertainties. Although the peak temperature ratio is slightly smaller than the integrated intensity ratio, the two values agree within the uncertainties. No significant gradients are seen in the $^{12}$CO/$^{13}$CO line ratio as a function of either space or velocity.

The $^{12}$CO/$^{13}$CO J=1-0 line ratio in MC20 agrees within the uncertainties with measurements of this line ratio in Galactic GMCs. For example, Solomon et al. (1979) obtained $R = 5.5 \pm 0.6$ from 17 lines of sight in the Galactic plane. The observed line ratios change by less than 40% between Galactic radii of 4 and 8 kpc and are consistent with a constant line ratio within the uncertainties. Polk et al. (1988) used complete $^{12}$CO and $^{13}$CO maps over 1.75 deg$^2$ of the Galactic plane to obtain $R = 4.5 \pm 0.6$ averaged over the entire area of GMCs and $R = 3.0 \pm 0.9$ at the emission peaks. Recently, Lee, Snell, & Dickman (1990) surveyed a 3 deg$^2$ region of the Galactic plane containing 47 GMCs. The average $^{12}$CO/$^{13}$CO peak temperature ratio in the GMCs in their map is $2.8 \pm 0.1$.

Because of the low signal-to-noise ratio of the $^{13}$CO single-dish spectra, using the integrated intensity measurements in Table 1 to determine the line ratio could introduce large uncertainties from the detailed shape of the baseline close to the line position. We chose instead to take the ratio of the two single-dish spectra and measure the average line ratio using only those channels for which the line ratio had a reasonable signal-to-noise level. The spectra of the line ratios and their signal-to-noise ratios are shown in Figure 2. We calculated the $^{12}$CO/$^{13}$CO line ratio from the single dish data by averaging the value of $R$ in all channels with $R/\sigma_R \geq 2$ (Table 1). This method gives $R = 10.0 \pm 0.5$ over 34 channels with an rms dispersion of 2.8. Including the uncertainty in the calibration of the single dish data (§2) gives $R = 10.0 \pm 0.9$. The average value of $R$ is not very sensitive to the signal-to-noise cutoff or the averaging method used.

The value of $R$ obtained from the single dish measurements of M33 is in good agreement with low-resolution measurements of other spiral galaxies, which give values ranging from 4 to 17 with an average value of about 11 (Encrenaz et al. 1979; Rickard & Blitz 1985; Young & Sanders 1986; Sage & Isbell 1991; Aalto et al. 1991). Similar values of $R$ are also obtained for early-type galaxies (Wiklind & Henkel 1990; Sage 1990), Markarian galaxies (Krügel, Steppe, & Chini 1990), and irregular galaxies (Becker 1990; Becker & Freudling 1991), while starburst and merging galaxies have $^{12}$CO/$^{13}$CO line ratios of 15 to 46 (Becker & Freudling 1991; Aalto et al. 1991, Combes et al. 1991, Casoli et al. 1991). However, the average single dish value of $R$ in M33 differs at the $5\sigma$ level from the value of $R$ measured in Galactic GMCs. We consider in the next section the various explanations which have been advanced to explain the high values of $R$ measured in external galaxies.

## 4. A Population of Diffuse Molecular Clouds in M33

Interferometer channel maps of the giant molecular cloud MC20 in the $^{12}$CO and $^{13}$CO lines are shown in Figure 3. The peak brightness temperature is 3.7 K ($9\sigma$) in the $^{12}$CO line and 0.64 K ($5\sigma$) in the $^{13}$CO line. Because of the different signal-to-noise ratios in the two lines, the maps are contoured as a percentage of the peak flux. When plotted in this manner, it is clear that MC20 has essentially the same appearance in the two lines as a function of velocity. The integrated intensity maps are shown in Figure 4, again contoured as a percentage of the peak, and the similar appearance of MC20 in these two lines is again striking. The cloud is resolved in both lines and the deconvolved diameters agree very well (Table 2). Integrated $^{12}$CO and $^{13}$CO spectra of MC20 (Figure 5) were obtained by integrating the emission in each channel map using a box with sides 1.4 times the observed full-width half-maximum diameter (see Wilson & Scoville 1990). Again, the velocity width (full-width half-maximum) is very similar in the two lines. The good agreement between the spatial and velocity extent of the $^{12}$CO and $^{13}$CO data indicates that the average filling factor of MC20 is the same in the two lines. Thus the large $^{12}$CO/$^{13}$CO line ratios observed in M33 (and probably in external galaxies in general) are unlikely to be caused by different filling factors in the two lines, as suggested by Encrenaz et al. (1979).

The single dish measurements of the $^{12}$CO/$^{13}$CO line ratio in M33 given in Table 1 do not show any significant variation as a function of position. Since the uncertainty in the $^{13}$CO data for a single channel is at best 15% (Table 1), the scatter of the seven line ratios about the mean value can probably be entirely explained by measurement uncertainties. Thus our detailed study of M33 confirms the results of Young & Sanders (1986) that the $^{12}$CO/$^{13}$CO line ratio does not vary between the centers and disks of galaxies. M33 has a significant abundance gradient (cf. Vilchez et al. 1988), ranging from $12 + log(O/H) = 9.0$ near the center to $12 + log(O/H) = 8.5$ in NGC 604, which is located at a radius of $\sim 3.5$ kpc. Sage & Isbell (1991) also found that the line ratio is independent of metallicity within a range $9.1 \leq 12 + log(O/H) \leq 9.6$. Regions with lower metallicities are expected to have higher [$^{12}$CO]/[$^{13}$CO] abundance ratios and hence line ratios due to the decreased importance of secondary processing (cf. Langer & Penzias 1990). Our results show that the line ratio is not sensitive to a factor of three change in the metallicity and so the high line ratios seen in the single dish data for M33 relative to those of Galactic GMCs cannot be attributed to abundance variations.

The remaining possible explanation for the high line ratios relies on the presence of diffuse molecular clouds. If M33 has a population of diffuse molecular clouds, they would contribute to the single dish line ratio, while the interferometric data would measure the line ratio in giant molecular clouds alone. There is evidence that the $^{12}$CO/$^{13}$CO J=1-0 line ratio is different in diffuse molecular clouds and GMCs. Blitz, Magnani, & Mundy (1984) obtained $R = 10.5$ in high latitude molecular clouds, while Knapp & Bowers (1988) obtained $R \geq 20$ for five very small ($< 1$ pc) molecular clouds. Indirect evidence also comes from recent analyses of large maps of the Galactic plane. The average line ratio in the Polk et al. (1988) data is $R = 6.7 \pm 0.7$, while the average line ratio within GMCs is $R = 4.5 \pm 0.6$. Adopting $R = 21$ for the diffuse molecular clouds, they estimate that 30-60% of the $^{12}$CO flux originates in diffuse molecular clouds. Similarly, the average line ratio in the map of Lee et al. (1990) is $R = 5.53 \pm 0.02$, while for emission from GMCs the line ratio is ~ 3; they estimate that at most 50% of the $^{12}$CO emission in this region could be due to diffuse molecular clouds. Having ruled out filling factors and metallicity, we conclude that the high single dish line ratios are evidence for a population of diffuse molecular clouds in the disk of M33.

### 5. How Important Are Diffuse Molecular Clouds in M33?

We have measured the $^{12}$CO/$^{13}$CO line ratio both within an individual molecular cloud and averaged over 200 pc diameter regions of M33. We can combine the two measured line ratios in M33 to probe the physical properties of the diffuse molecular clouds, which are too small and faint to be studied directly. Let $F$ be the ratio of the $^{12}$CO flux from diffuse molecular clouds to the $^{12}$CO flux from giant molecular clouds, $R_{GMC}$ be the intrinsic value of $R$ in giant molecular clouds, $R_{diff}$ be the intrinsic value of $R$ in diffuse molecular clouds, and $R_{NRAO}$ be the average value of $R$ observed with the NRAO 12 m telescope. These four quantities are then related by the equation

$$F = \frac{R_{diff}}{R_{GMC}} \frac{R_{NRAO} - R_{GMC}}{R_{diff} - R_{NRAO}} \qquad (1)$$

(equation (3) from Polk et al. 1988, adapted to our notation). We can solve this equation for $R_{diff}$, which gives

$$R_{diff} = \frac{F R_{NRAO} R_{GMC}}{(1+F) R_{GMC} - R_{NRAO}} \qquad (2)$$

If both the $^{12}$CO and the $^{13}$CO emission were optically thin, the observed line ratio would be equal to the abundance ratio. Thus the upper limit to $R_{diff}$ is the [$^{12}$CO]/[$^{13}$CO] abundance ratio; adopting the maximum value for $R_{diff}$ yields a lower limit to $F$. Comparing the total flux detected in the interferometer observations of giant molecular clouds with the single-dish flux at the same position and assuming that all the flux missed by the interferometer is contained in diffuse molecular clouds (rather than small giant molecular clouds) gives an upper limit to $F$. Adopting this upper limit to $F$ in equation (2) yields a lower limit to $R_{diff}$. Thus we can use our observations of the $^{12}$CO/$^{13}$CO line ratio to set a lower limit on both the fraction of the $^{12}$CO emission in M33 that originates in diffuse molecular clouds and the $^{12}$CO/$^{13}$CO line ratio in the diffuse molecular clouds.

From our data we have $R_{NRAO} = 10.0 \pm 0.9$ (§3). Since we only have a single set of interferometric $^{13}$CO measurements, we must adopt the observed line ratio in MC20 as representative of the average value, $R_{GMC} = 7.5 \pm 2.1$ (§3). We adopt $50 \pm 20$ as the [$^{12}$CO]/[$^{13}$CO] abundance ratio and hence the upper limit to $R_{diff}$. These values give a lower limit to $F$ of $0.4 \pm 0.5$. We have used the published $^{12}$CO fluxes (Wilson & Scoville 1990) for molecular clouds that lie in our single dish beams to calculate an upper limit to $F$ for each position (Table 3). Excluding the unusual region MC1-5, the average upper limit is $F \leq 1.4 \pm 0.2$. This analysis indicates that between ~30% and 60% of the total $^{12}$CO emission from M33 may be due to a population of diffuse molecular clouds. The value of 60% is a firm upper limit to the diffuse cloud emission fraction, since the remainder of the emission is seen directly to originate in giant molecular clouds. Due to the large uncertainties in the value of $R$ measured for MC20, the lower limit of 30% is more uncertain. If giant molecular clouds in M33 have the same value of $R_{GMC}$ as Galactic

GMCs (5 ± 1), then we obtain $F = 1.3 ± 0.9$ and the diffuse clouds contribute $60^{+10}_{-30}\%$ of the total $^{12}$CO emission, in good agreement with the upper limit estimated from the interferometric data.

Adopting 1.4 as an upper limit to $F$ gives a lower limit to the $^{12}$CO/$^{13}$CO line ratio in the diffuse molecular clouds of $R_{diff} \geq 13 ± 5$. Thus the $^{12}$CO/$^{13}$CO line ratio in any diffuse molecular clouds in M33 is somewhat larger than the value measured in the giant molecular cloud MC20. The dominant contribution to the uncertainty in $R_{diff}$ comes from the uncertainty in $R_{GMC}$, which is large because it is based on a single measurement and so includes the uncertainty in the absolute calibration. In addition, the absolute calibration of the older $^{12}$CO data is not as good as the more recent data, since the receivers were not as sensitive in 1987 as they are today. $^{13}$CO interferometer maps of additional giant molecular clouds in M33 would probably reduce the uncertainty in the average value of $R_{GMC}$, since the calibration uncertainty would be independent from one cloud to another. Reducing the uncertainty in $R_{GMC}$ by a factor of two would reduce the uncertainty in $R_{diff}$ from ±5.3 to ±3.6, while reducing it by a factor of three (to ∼10% if $R_{GMC} \sim 7$) would reduce the uncertainty in $R_{diff}$ to ±3.2. Thus additional $^{13}$CO interferometric observations could be quite successful in separating the emission properties of diffuse molecular clouds from those of giant molecular clouds. At present, since the three line ratios (GMCs, diffuse molecular clouds, and disk-averaged) all agree within the uncertainties, the conclusion that M33 contains a population of diffuse molecular clouds with different CO emission properties from giant molecular clouds rests primarily on the difference between the average single dish line ratio and the ratio obtained for Galactic GMCs.

## 6. The Physical Conditions in the Molecular Gas in M33
### 6.1. The Average Properties of the Molecular Gas

We can combine the $^{12}$CO/$^{13}$CO J=1-0 single dish line ratios measured here with recent measurements of the $^{12}$CO J=2-1 line in M33 at the same resolution (Thornley & Wilson 1994) to try to constrain the physical conditions in the molecular gas. We will use the Large Velocity Gradient (LVG) approximation to model the line ratios. To compare our observed line ratios with the peak brightness temperatures produced by the models, we assume that the filling factor of the CO lines is the same for each transition and isotope. (The interferometer results for MC20 show that the area covered by this cloud is the same in both the $^{12}$CO and the $^{13}$CO J=1-0 lines. However, we cannot determine the filling factor of the diffuse molecular clouds (see Thornley & Wilson 1994).) The average dust temperature in M33 deduced from IRAS data is ∼ 30 K over a wide range of galactic radius (Rice et al. 1990). This value is likely an upper limit to the average dust temperature since the IRAS data are not sensitive to emission from cold dust. The [$^{12}$CO]/[$^{13}$CO] abundance ratio in the Galaxy ranges from a value of ∼ 30 in the Galactic Center to a value of ∼ 70 at a radius of 12 kpc (Langer & Penzias 1990). In our models we use kinetic temperatures ranging from 10 to 30 K and [$^{12}$CO]/[$^{13}$CO] abundance ratios ranging from 30 to 70. The molecular hydrogen density was varied between $10^2$ and $10^4$ cm$^{-3}$ and the $^{12}$CO column density per unit velocity ($N(^{12}CO)/dv$) ranged from $10^{16}$ to $10^{20}$ cm$^{-2}$ (km s$^{-1}$)$^{-1}$.

The two CO line ratios ($^{12}$CO(J=1-0)/$^{13}$CO(J=1-0) and $^{12}$CO(J=2-1)/$^{12}$CO(J=1-0)) are plotted as a function of density and column density for four different kinetic temperatures in Figure 6. The observed line ratios constrain the column densities to within an order of magnitude, but only weak constraints are placed on the density at a given temperature. It is easy to understand the physical basis of this result. The $^{13}$CO line is most likely optically thin, and so the predicted brightness temperature is proportional to the column density of $^{13}$CO. The uncertainty in the $^{12}$CO column density at a given temperature is mainly due to the possible range in the [$^{12}$CO]/[$^{13}$CO] abundance ratio. For the full range of abundance ratios and temperatures, the $^{12}$CO column density is constrained to lie in the range $2 \times 10^{16} - 3 \times 10^{17}$ cm$^{-2}$ (km s$^{-1}$)$^{-1}$, while $^{13}$CO column density must be between $5 \times 10^{14}$ and $4 \times 10^{15}$ cm$^{-2}$ (km s$^{-1}$)$^{-1}$. For a single temperature the range of allowed $^{13}$CO column density is only a factor of four. However, without measurements of additional line ratios, we cannot determine the density or the kinetic temperature of the gas.

### 6.2. The Conditions in Diffuse and Giant Molecular Clouds

What do the different $^{12}$CO/$^{13}$CO J=1-0 line ratios determined for giant and diffuse molecular clouds imply for the physical conditions in these clouds? Since the average $^{12}$CO line ratio measured by Thornley & Wilson (1994) agrees within the uncertainties with the average values measured for translucent and high latitude clouds by van Dishoeck et al. (1991), we will assume that the $^{12}$CO J=2-1 to 1-0 line ratio is the same in each type of cloud. If the two types of clouds have the same intrinsic density and temperature, the diffuse molecular clouds would need a larger [$^{12}$CO]/[$^{13}$CO] abundance ratio than the GMCs to produce the larger $^{12}$CO/$^{13}$CO J=1-0 line ratio. However, fractionation tends to reduce the [$^{12}$CO]/[$^{13}$CO] abundance ratio below the intrinsic [$^{12}$C]/[$^{13}$C] abundance ratio in regions with a large abundance of C$^+$. Fractionation should be more important in diffuse molecular clouds because of their lower total column density, and so we would expect that diffuse molecular clouds would have a *lower* [$^{12}$CO]/[$^{13}$CO] abundance ratio than giant molecular clouds. Therefore, it is unlikely that the diffuse and giant molecular clouds have the same density and kinetic temperature.

If the diffuse and giant molecular clouds have the same kinetic temperature and [$^{12}$CO]/[$^{13}$CO] abundance ratio, then the diffuse molecular clouds must have higher densities than the GMCs. Allowing the diffuse molecular clouds to have a lower [$^{12}$CO]/[$^{13}$CO] abundance ratio than GMCs further increases the density difference. On the other hand, translucent molecular clouds in the Galaxy are observed to have kinetic temperatures ranging from 15-50 K (van Dishoeck et al. 1991, and references therein). If the diffuse molecular clouds have a higher kinetic temperature than the GMCs, then they must have a lower average density. Obviously more data are required to determine any temperature or density differences between diffuse and giant molecular clouds.

If there are two populations of molecular clouds in M33 with different physical properties, the use of the CO-to-H$_2$ conversion factor obtained from giant molecular clouds could introduce systematic errors in the calculation of the global gas mass. How large are such errors likely to be? The diffuse clouds can contribute up to 60% of the observed $^{12}$CO J=1-0 emission from M33 (Wilson & Scoville 1990; §5). For a population of virialized molecular clouds in an external galaxy, the CO-to-H$_2$ conversion factor is expected to vary as $\sqrt{n}/T_B$ (Dickman, Snell, & Schloerb 1986), where $T_B$ is the intrinsic brightness temperature of the clouds. If the diffuse and giant molecular clouds have the same kinetic temperature and abundance ratio, the diffuse clouds would be three times denser than the GMCs and the global molecular gas mass would be underestimated by $\sim 40\%$. However, the diffuse molecular clouds may not be gravitationally bound. Alternatively, if we assume that the CO-to-H$_2$ conversion factor for the M33 diffuse molecular clouds is the same as that of diffuse clouds in our Galaxy ($0.5 \times 10^{20}$ cm$^{-2}$ (K km s$^{-1}$)$^{-1}$, de Vries, Heithausen, & Thaddeus 1987), then the global molecular gas mass would be overestimated by about 50%. Thus the effect on the global gas mass depends critically on the precise physical conditions in the diffuse molecular clouds, but the H$_2$ gas masses in M33 are unlikely to be incorrect by more than a factor of two.

## 7. Conclusions

We have obtained $^{12}$CO and $^{13}$CO J=1-0 observations of the nearby spiral galaxy M33 to try to resolve the discrepancy between measurements of the $^{12}$CO/$^{13}$CO line ratio in our own and other galaxies, as well as to study the physical conditions in the molecular interstellar medium and to look for evidence of a population of diffuse molecular clouds. The main results are summarized below.

(1) The $^{12}$CO/$^{13}$CO line ratio obtained from single dish observations centered on seven giant molecular clouds is $10.0 \pm 0.9$. This value is in good agreement with line ratios obtained in other normal galaxies, but is significantly larger than the line ratio measured in Galactic Galactic giant molecular clouds. The interferometer maps of the giant molecular cloud MC20 show that the $^{12}$CO/$^{13}$CO integrated intensity ratio in this cloud is $7.5 \pm 2.1$. This line ratio agrees within the uncertainties with line ratios measured for GMCs.

(2) Three explanations have been put forward previously to explain the different $^{12}$CO/$^{13}$CO line ratios observed in Galactic GMCs and external galaxies: different filling factors in the two lines; abundance variations; and the presence of diffuse molecular clouds. The interferometer maps of MC20 show that the cloud has a similar spatial and velocity extent in the two CO lines and thus the filling factors of the two lines are the same. In addition, we see no evidence for significant variations in the line ratio from one position to another in M33. One of our observed positions is centered on the giant HII region NGC 604, which has a metallicity that is lower by a factor of three than the inner disk regions, and thus small changes in the metallicity do not affect the line ratio. We conclude that the contribution of diffuse molecular clouds to the measured single dish lines is the most likely explanation for the high $^{12}$CO/$^{13}$CO line ratios observed in M33, and in the disks of spiral galaxies in general.

(3) We have combined our single dish and interferometric measurements of the CO line ratio to obtain a lower limit of $13 \pm 5$ for the $^{12}$CO/$^{13}$CO line ratio in the diffuse molecular clouds in M33. In addition, we obtain a lower limit of $30 \pm 30\%$ for the fraction of the total $^{12}$CO emission from M33 that originates in diffuse molecular clouds. An upper limit of $\sim 60\%$ is obtained by comparing the amount of flux detected in the interferometric and the single dish data. We estimate that the effect of the different physical conditions in diffuse and giant molecular clouds in M33 is unlikely to cause global molecular gas masses to be in error by more than a factor of two.

(4) We have combined our $^{12}$CO/$^{13}$CO line ratio with the $^{12}$CO J=2-1 to J=1-0 line ratio measured by Thornley & Wilson (1994) to constrain the physical conditions in the molecular interstellar medium using large velocity gradient models. The $^{12}$CO and $^{13}$CO column densities can be determined to within an order of magnitude, while the density and temperature of the gas are still not well constrained by these two line ratios. The different $^{12}$CO/$^{13}$CO line ratios measured in the diffuse and giant molecular clouds in M33 suggest only that the two types of clouds are unlikely to have the same kinetic temperature and average density. If the density and temperature were the same, then the diffuse clouds would need to have a higher [$^{12}$CO]/[$^{13}$CO] abundance ratio than the giant molecular clouds, which is contrary to the expected effect of fractionation in a lower column density environment.

## Acknowledgements

We would like to thank the staff at NRAO and OVRO for their able assistance during our observing runs. We also thank Lee Mundy for the use of his LVG code. We thank the referee, Jill Knapp, for comments that greatly improved the presentation of this paper. C.D.W. was partially supported by NSF grant AST 91-00306, by funds from the Margaret Cullinan Wray Charitable Lead Annuity Trust, and by NSERC Canada through a Women's Faculty Award. C.E.W. was supported by the Keck Foundation.

## References

Aalto, S., Black, J. H., Johansson, L. E. B., & Booth, R. S. 1991, A&A, 249, 323
Becker, R. 1990, Ph. D. thesis, U. Bonn
Becker, R., & Freudling, W. 1991, A&A, 251, 454
Blitz, L., Magnani, L., & Mundy, L. 1984, ApJ, 282, L9
Casoli, F., Dupraz, C., & Combes, F. 1992, A&A, 264, 55
Casoli, F., Dupraz, C., Combes, F., & Kazès, I. 1991, A&A, 251, 1
Combes, F., Casoli, F., Encrenaz, P., Gerin, M., & Laurent, C. 1991, A&A, 248, 607
de Vries, H. W., Heithausen, A., & Thaddeus, P., 1987, ApJ, 319, 723
Dickman, R. L., Snell, R. L., & Schloerb, F. P. 1986, ApJ, 309, 326
Encrenaz, P. J., Stark, A. A., Combes, F., & Wilson, R. W. 1979, A&A, 78, 1
Gordon, M. A., & Burton, W. B. 1976, ApJ, 208, 346
Knapp, G. R. & Bowers, P. F. 1988, ApJ, 331, 974
Krügel, E., Steppe, H., & Chini, R. 1990, A&A, 229, 17
Langer, W. D., & Penzias, A. A. 1990, ApJ, 357, 477
Lee, Y., Snell, R. L., & Dickman R. L. 1990, ApJ, 355, 536
Polk, K. S., Knapp, G. R., Stark, A. A., & Wilson, R. W. 1988, ApJ, 332, 432
Rice, W., Boulanger, F., Viallefond, F., Soifer, B. T., & Freedman, W. L. 1990, ApJ, 358, 418
Rickard, L. J., & Blitz, L. 1985, ApJ, 292, L57
Rubio, M. et al. 1993, A&A, 271, 1
Sage, L. J. 1990, A&A, 239, 125
Sage, L. J., & Isbell, D. W. 1991, A&A 247, 320
Scoville, N. Z., & Sanders, D. B. 1987, in Interstellar Processes, es. D. J. Hollenbach & H. A. Thronson, [Reidel: Dordrecht], 21
Solomon, P. M., Scoville, N. Z., & Sanders, D. B. 1979, ApJ, 232, L89
Strong, A. W., et al. 1986, A&A, 207, 1
Thornley, M. D., & Wilson, C. D. 1994, ApJ, in press
van den Bergh, S. 1991, PASP, 103, 609
van Dishoeck, E. F., Black, J. H., Phillips, T. G., & Gredel, R. 1991, ApJ, 366, 141
Vilchez, J. M., Pagel, B. E. J., Diaz, A. I., Terlevich, E., & Edmunds, M. G. 1988, MNRAS, 235, 633
Wiklind, T., & Henkel, C. 1990, A&A, 227, 394
Wilson, C. D., & Scoville, N. 1990, ApJ, 363, 435
Wilson, C. D., & Scoville, N. 1992, ApJ, 385, 512
Young, J. S., & Sanders, D. B. 1986, ApJ, 302, 680

Table 1

$^{12}$CO and $^{13}$CO NRAO 12m Observations of M33 Centered on Molecular Clouds

| Position Name | $\alpha$(1950) ($^h$ $^m$ $^s$) | $\delta$(1950) ($^o$ $^\prime$ $^{\prime\prime}$) | $S(^{12}CO)$ (K km s$^{-1}$) | $T(^{12}CO)^a$ (K) | $S(^{13}CO)$ (K km s$^{-1}$) | $T(^{13}CO)^a$ (mK) | $R^b$ |
|---|---|---|---|---|---|---|---|
| MC1-5 | 01 31 03.0 | 30 23 54 | 1.61±0.24 | 0.12 ± 0.02 | 0.22±0.06 | 13 ± 4 | 9.2±2.0 |
| MC6-10 | 01 31 03.0 | 30 21 54 | 3.20±0.18 | 0.19 ± 0.01 | 0.25±0.06 | 14 ± 4 | 13.4±1.3 |
| MC13 | 01 31 10.3 | 30 20 40 | 2.22±0.20 | 0.16 ± 0.02 | 0.15±0.07 | 19 ± 4 | 10.1±0.6 |
| MC19 | 01 31 13.4 | 30 23 52 | 2.32±0.29 | 0.20 ± 0.02 | 0.16±0.06 | 19 ± 4 | 11.3±0.8 |
| MC20 | 01 31 10.9 | 30 25 24 | 3.10±0.21 | 0.24 ± 0.02 | 0.35±0.06 | 26 ± 5 | 9.0±0.4 |
| MC32 | 01 30 51.8 | 30 23 52 | 2.89±0.29 | 0.19 ± 0.03 | 0.25±0.06 | 26 ± 4 | 8.9±0.9 |
| NGC 604$^c$ | 01 31 44.5 | 30 31 15 | 4.43±0.26 | 0.18 ± 0.02 | 0.35±0.07 | 21 ± 4 | 8.5±0.9 |
| -220 km s$^{-1}$ | " | " | 1.94±0.18 | 0.18 ± 0.02 | 0.21±0.05 | 21 ± 4 | 7.4±0.6 |
| -240 km s$^{-1}$ | " | " | 2.49±0.18 | 0.17 ± 0.02 | 0.14±0.05 | 10 ± 4 | 11.9±2.0 |

$^a$ Calculated in 2.6 km s$^{-1}$ channels; uncertainty is the rms noise in the spectrum.
$^b$ $^{12}$CO/$^{13}$CO line ratio, calculated using only channels with a signal-to-noise in the line ratio greater than 2. The uncertainty in the absolute calibration is not included.
$^c$ Data for the two separate velocity components also given.

# Table 2
## Properties of the Molecular Cloud MC20

| Property | $^{12}CO^a$ | $^{13}CO$ |
|---|---|---|
| Deconvolved Diameter[b] (FWHM) (pc) | 29x34 | 29x38 |
| Peak Temperature (K) | $3.7 \pm 0.4$ | $0.64 \pm 0.11$ |
| Integrated Intensity[c] (Jy km s$^{-1}$) | $36.7 \pm 2.7$ | $4.9 \pm 0.8$ |
| Velocity Width (FWHM) (km s$^{-1}$) | 9.2 | 8.7 |
| Optical Depth[d] | 7 | 0.14 |
| Column Density[d,e] (cm$^{-2}$) | $2.8 \times 10^{11}$ | $5.7 \times 10^{9}$ |
| Virial Mass[f] (M$_\odot$) | $3.7 \times 10^5$ | $3.5 \times 10^5$ |
| Molecular Mass[g,h] (M$_\odot$) | $3.7 \times 10^5$ | $3.3 \times 10^5$ |

A distance to M33 of 0.79 Mpc is assumed throughout (van den Bergh 1991).

[a] $^{12}CO$ properties differ slightly from those of Wilson & Scoville, due to the use of uniform weighting and the exclusion of faint $^{12}CO$ wings beyond the $^{13}CO$ noise limit.

[b] Beam size is 30x23 pc.

[c] Integrated over 15.6 km s$^{-1}$; the $^{12}CO$ emission has broader faint wings.

[d] Assumes [$^{12}CO$]/[$^{13}CO$] = 50.

[e] $N(^{13}CO) = 8.06 \times 10^8 (1 - e^{-5.28/T_{ex}})^{-1} T_{ex} \tau$ cm$^{-2}$; assumes optically thin emission and T$_{ex}$=15 K.

[f] $M_{vir} = 99 \Delta V_{FWHM}^2 \overline{D}$, where $\overline{D} = 1.4 \times \overline{D_{FWHM}}$; assumes a $1/r$ density profile.

[g] For $^{12}CO$, $M_{mol} = 1.61 \times 10^4 d^2_{Mpc} S_{12}$ (assumes $\alpha = 3 \times 10^{20}$ cm$^{-2}$ (K km s$^{-1}$)$^{-1}$, Strong et al. 1986, Scoville & Sanders 1987);

[h] for $^{13}CO$, $M_{mol} = 1.71 \times 10^{-15} f D_{pc}^2 \Delta V_{FWZI} ([^{13}CO]/[H_2])^{-1} N(^{13}CO)$; assumes a filling factor $f = 1$, [$^{13}CO$]/[H$_2$]=10$^{-6}$, and $\Delta V_{FWZI} = 15.6$ km s$^{-1}$.

Table 3

**The $^{12}$CO Flux Detected by OVRO and NRAO in M33**

| Position Name | NRAO Flux$^a$ (Jy km s$^{-1}$) | OVRO Flux$^b$ (Jy km s$^{-1}$) | Fraction Detected with OVRO$^c$ | F$^d$ |
|---|---|---|---|---|
| MC1-5 | 54.7 | 60 | 1.1 | 0 |
| MC6-10 | 108.8 | 50 | 0.5 | 1.2 |
| MC13 | 74.8 | 22 | 0.3 | 2.4 |
| MC19 | 78.2 | 27 | 0.3 | 1.9 |
| MC20 | 105.4 | 52 | 0.5 | 1.0 |
| MC32 | 102.0 | 46 | 0.5 | 1.2 |
| NGC 604 | 150.6 | 85 | 0.6 | 0.8 |

$^a$ From Table 1; assumes 34 Jy K(T$^*_R$)$^{-1}$.
$^b$ Calculated from data in Wilson & Scoville (1990); includes the flux of all clouds within 27$''$ of the pointing center, weighted by a 54$''$ full-width half-maximum Gaussian beam.
$^c$ OVRO flux divided by NRAO flux.
$^d$ $F = (S(NRAO) - S(OVRO))/S(OVRO)$

# Figure Captions

**Fig. 1:** $^{12}$CO and $^{13}$CO J=1-0 spectra obtained at the NRAO 12 m telescope for seven positions in M33 centered on giant molecular clouds. The beam size is 55″ or 210 pc (FWHM), much larger than the diameter of the molecular cloud. The $^{13}$CO spectra have been multiplied by a factor of 10. The molecular cloud identifications are from Wilson & Scoville (1990).

**Fig. 2:** The $^{12}$CO/$^{13}$CO line ratio as a function of velocity for the seven positions in M33 and the signal to noise in the line ratio.

**Fig. 3:** Channel maps of the giant molecular cloud MC20 in $^{12}$CO and $^{13}$CO obtained with the Owens Valley Millimeter-Wave Interferometer. Only channels containing emission at greater than the $2\sigma$ level in $^{13}$CO are shown. The synthesized beam is $7.8'' \times 6.1''$. The maps have been corrected for attenuation due to the primary beam. (a) $^{12}$CO emission. The noise in the individual channels is 0.19 Jy/beam and the peak emission is 1.90 Jy/beam ($10\sigma$). The contours are $(-1, 1, 2, 3, 4, 5) \times 20\%$ of the peak and negative emission is shown by dashed contours. (b) $^{13}$CO emission. The noise in the individual channels is 0.058 Jy/beam and the peak emission is 0.29 Jy/beam ($5\sigma$). The contours are $(1, 2, 3, 4, 5) \times 20\%$ of the peak; the contour at -20% is omitted for clarity.

**Fig. 4:** Integrated intensity maps of MC20 in the $^{12}$CO and $^{13}$CO emission lines. The synthesized beam is $7.8'' \times 6.1''$. Contour levels are $(-1, 1, 2, 3, 4, 5) \times 20\%$ of the peak. The peak emission in the $^{12}$CO map is 1.10 Jy/beam ($11\sigma$), while the peak emission in the $^{13}$CO map is 0.16 Jy/beam ($5.9\sigma$). The maps are integrated over the 6 channels shown in Figure 3 (15.6 km s$^{-1}$).

**Fig. 5:** Spectra of the $^{12}$CO and $^{13}$CO emission from MC20 obtained from the OVRO channel maps. The flux has been integrated over the area of the cloud. The $^{13}$CO spectrum has been multiplied by a factor of 7.5.

**Fig. 6:** Two CO line ratios are plotted as a function of density and column density for four different kinetic temperatures (10, 15, 20, 30K). Dashed line: the $^{12}$CO J=2-1 to J=1-0 ratio, $0.67 \pm 0.19$ (Thornley & Wilson 1994). The highest contour (0.86) is off the upper edge of the $T_K = 10$ K plot. Solid line: the $^{12}$CO/$^{13}$CO J=1-0 ratio, 10.0, plotted for three values of the [$^{12}$CO]/[$^{13}$CO] abundance, 30, 50, and 70.

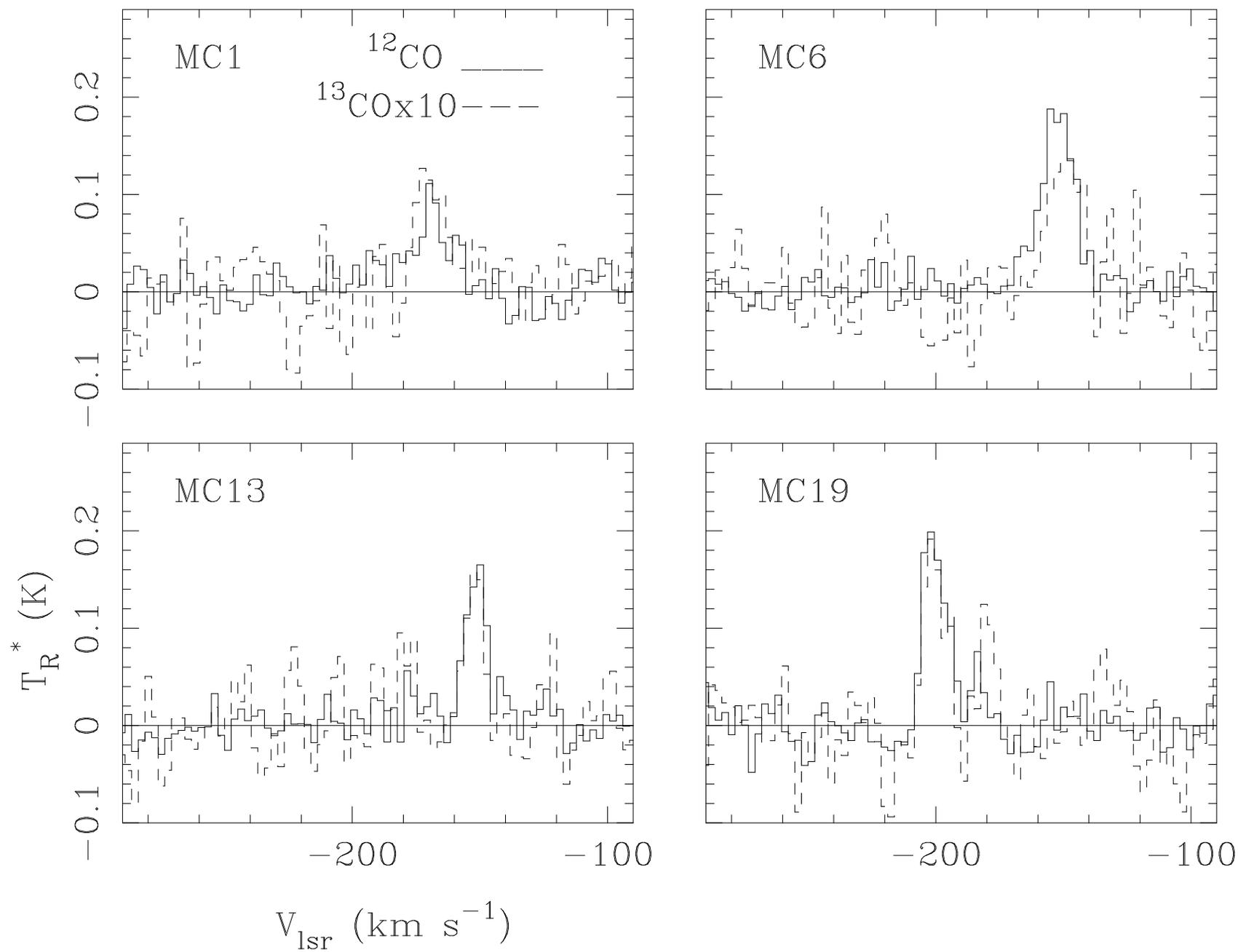

Figure 1

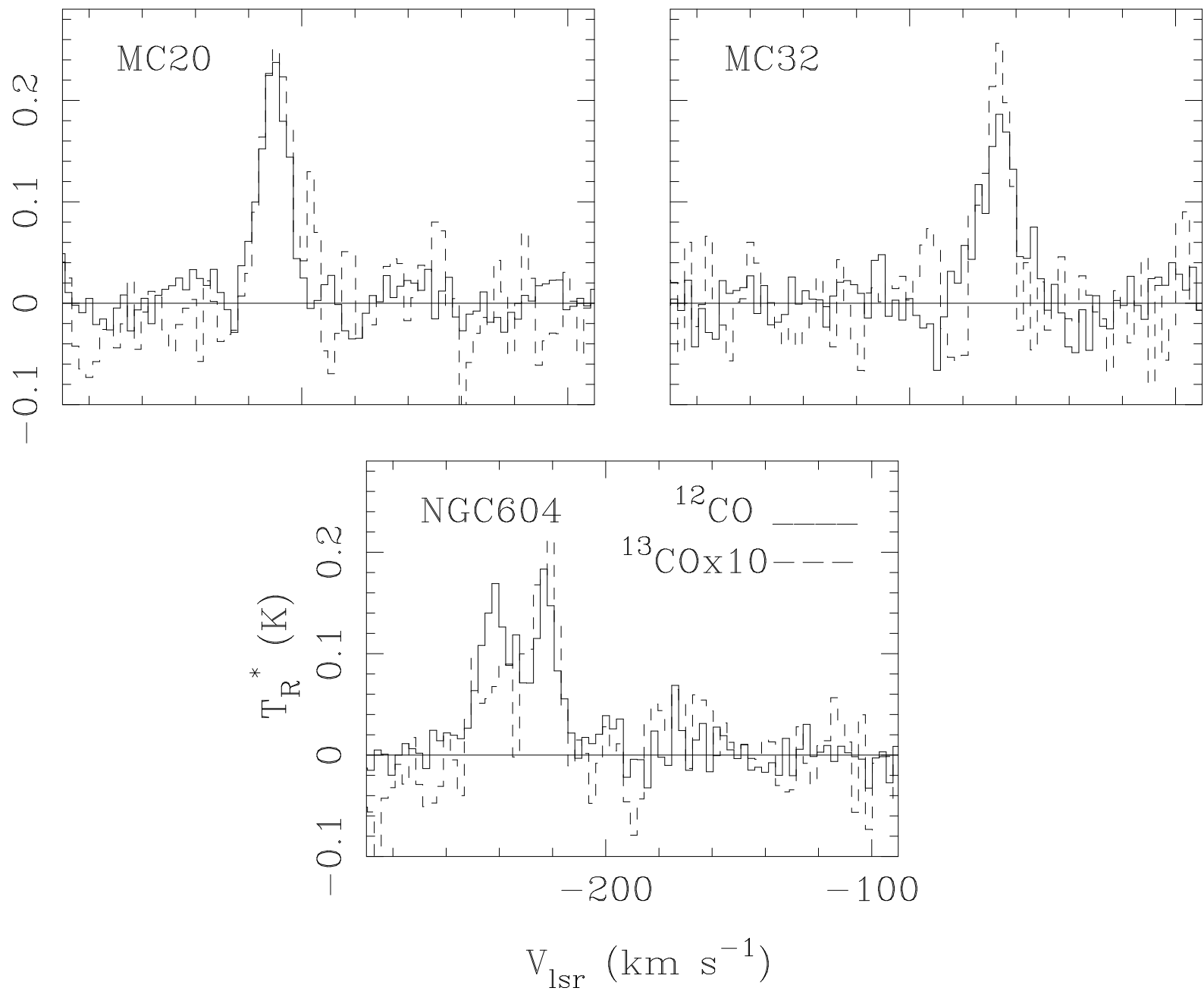

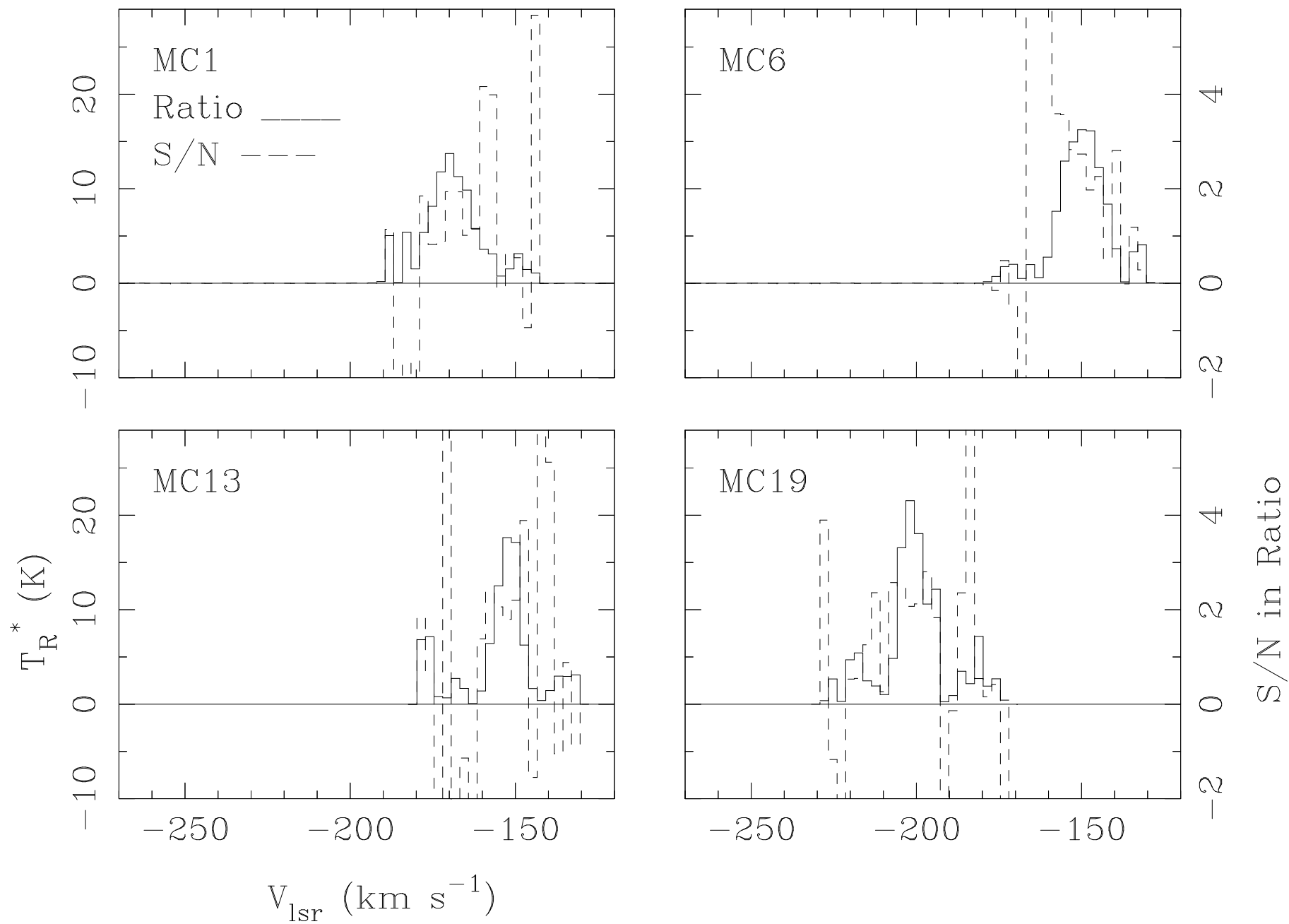

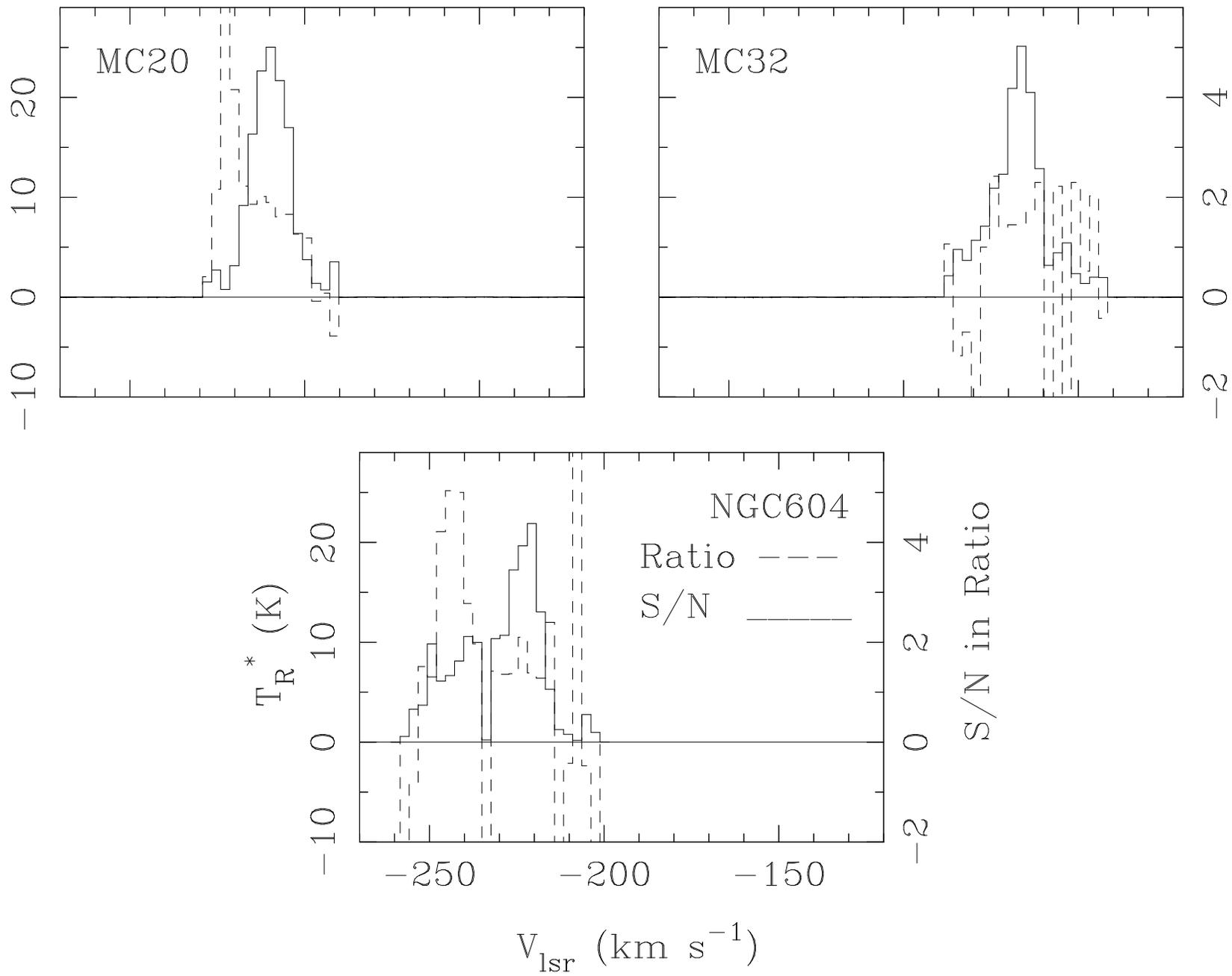

Figure 3a

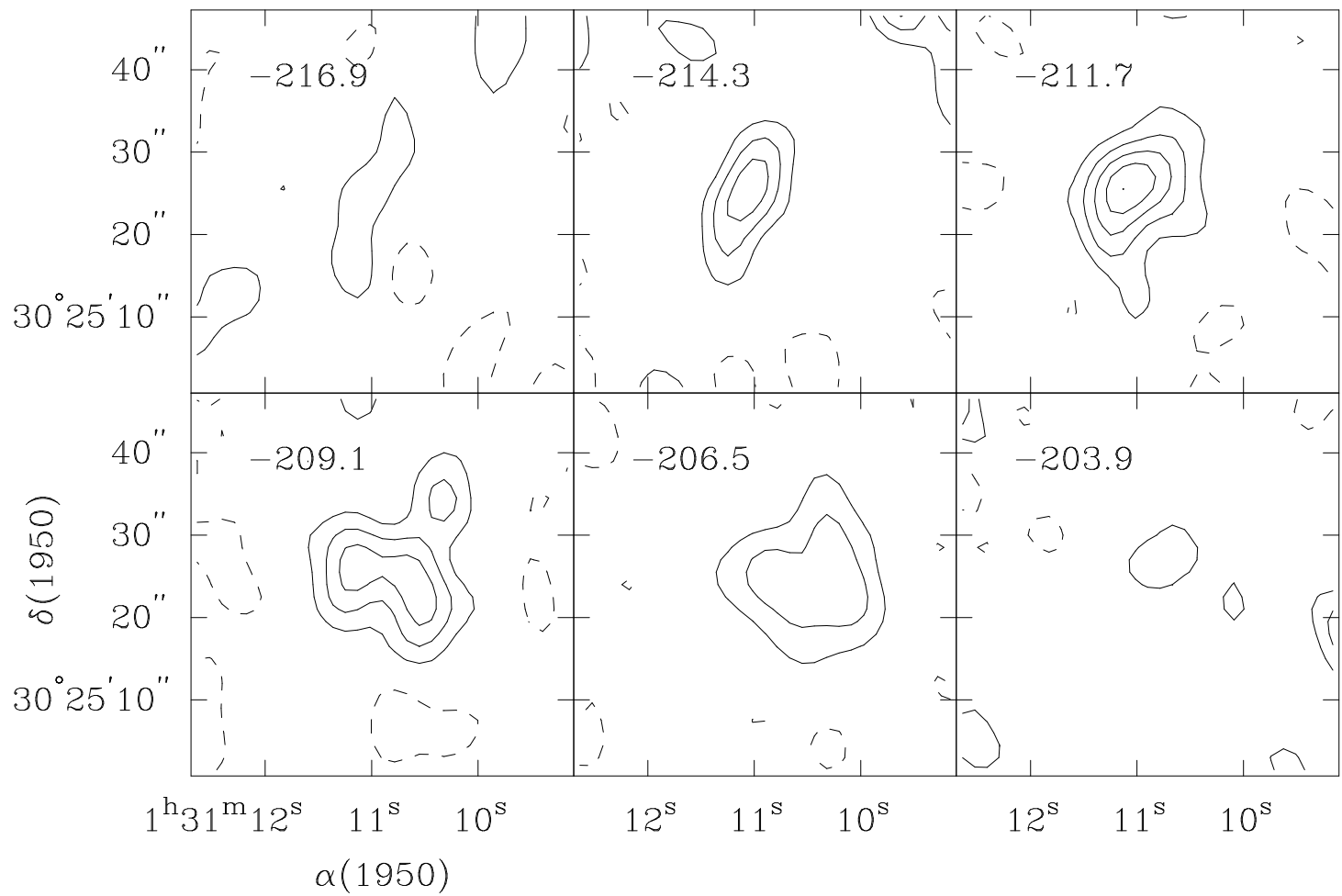

Figure 3b

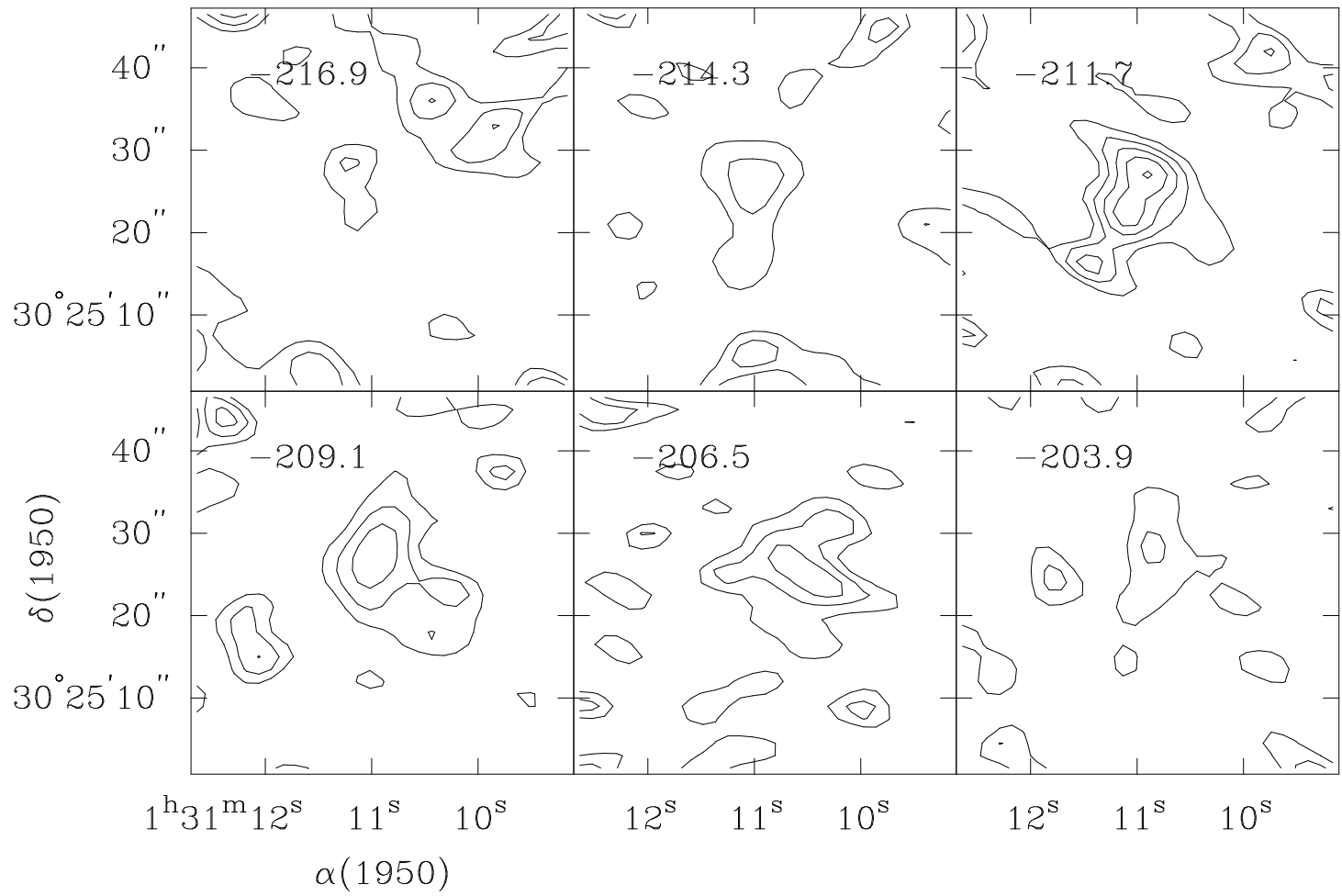

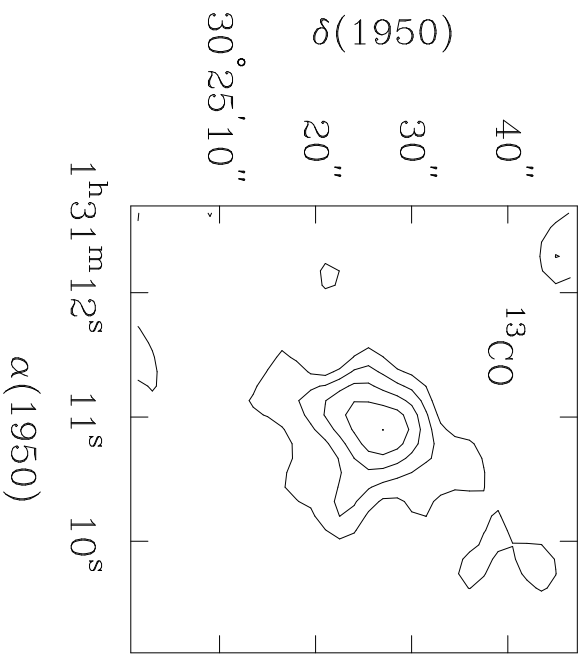
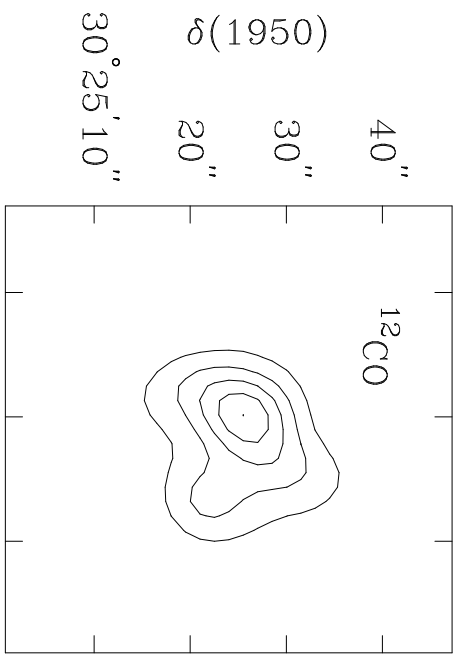

Figure 4

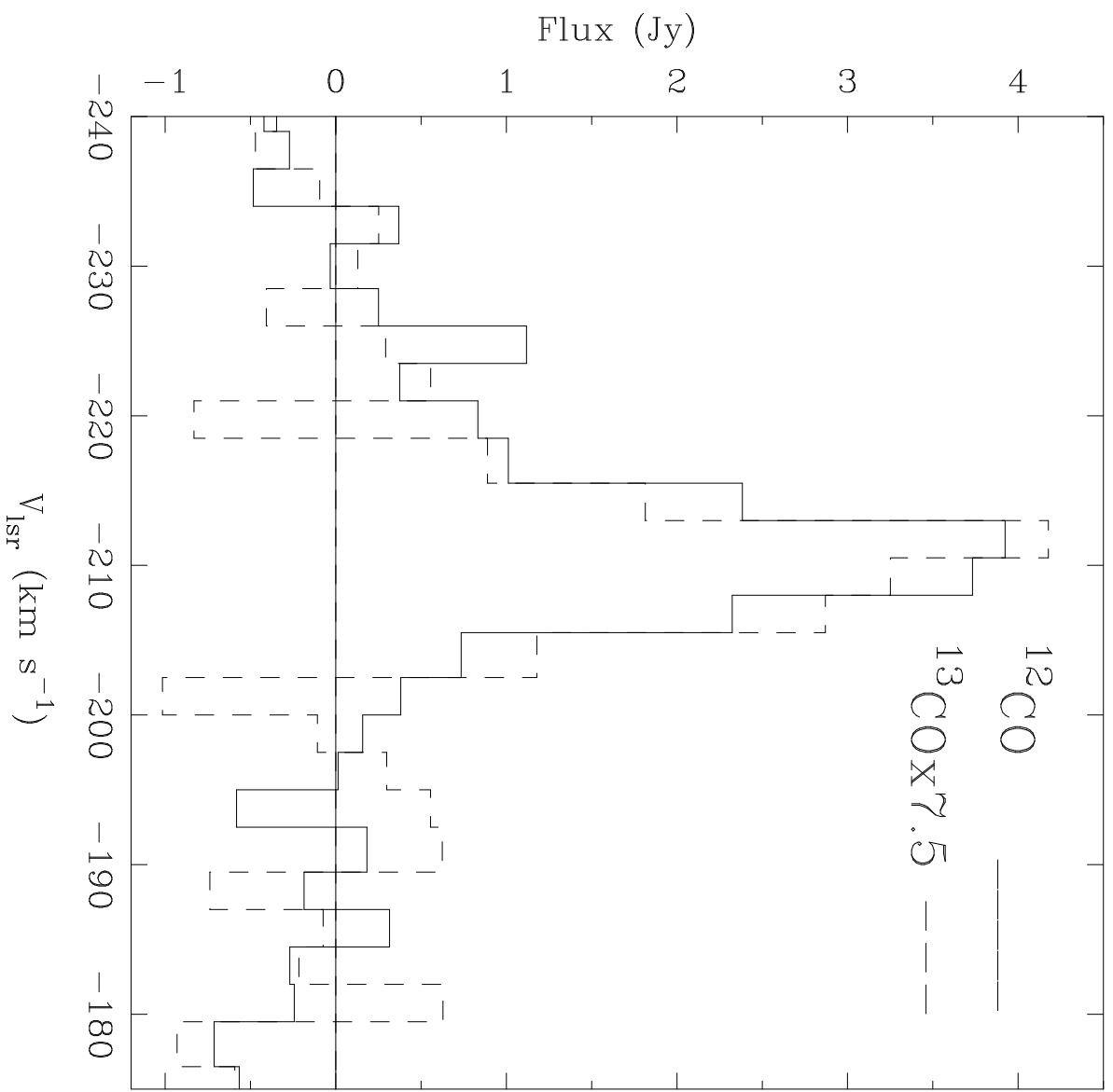

Figure 5

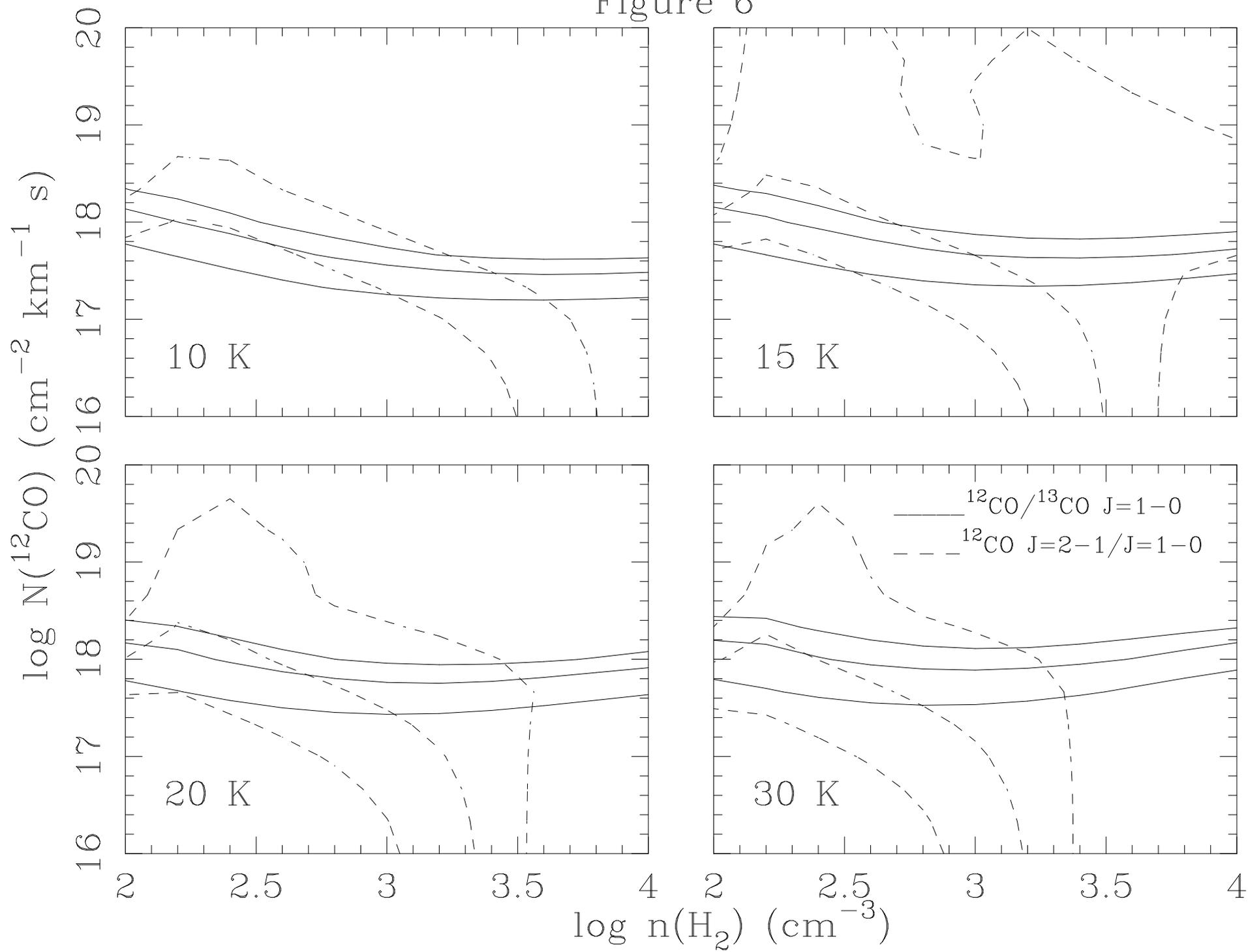

Figure 6